\input{aipcheck}

\documentclass[
    ,final            % use final for the camera ready runs
  ]
  {aipproc}

\layoutstyle{6x9}

\begin{document}

\title{Gamma-Ray Bursts and Afterglow Polarisation}


\author{S. Covino}{
address={INAF / Brera Astronomical Observatory, V. Bianchi 46, 22055, Merate (LC), Italy}
}

\author{E. Rossi}{
address={Max Planck Institute for Astrophysics, Garching,
Karl-Schwarzschild-Str. 1, 85741 Garching, Germany}
}

\author{D. Lazzati}{
address={JILA, University of Colorado, 440 UCB, Boulder, CO 80309-0440, USA}
}

\author{D. Malesani}{
address={International School for Advanced Studies (SISSA-ISAS), via Beirut 2-4, I-34014 Trieste, Italy}
}

\author{G. Ghisellini}{
address={INAF / Brera Astronomical Observatory, V. Bianchi 46, 22055,
Merate (LC), Italy} }

\begin{abstract}
Polarimetry of Gamma-Ray Burst (GRB) afterglows in the last few years
has been considered one of the most effective tool to probe the
geometry, energetic, dynamics and the environment of GRBs. We report
some of the most recent results and discuss their implications and
future perspectives.
\end{abstract}

\maketitle

%%%%%%%%%%%%%%%%%%%%%%%%%%%%%%%%%%%%%%%%%%%%
%% MAINMATTER
%%%%%%%%%%%%%%%%%%%%%%%%%%%%%%%%%%%%%%%%%%%%

\section{Introduction}

Polarimetry has always been a niche observational technique. It may be
difficult to apply, requiring special care for the instruments, data
reduction and analysis. Indeed, for real astronomical sources, where
often the polarisation degree is fairly small at the level of a few
per cent, the signal to noise required to derive useful information
has to be very high.  However, the amount of information that can be
extracted by a polarised flux is also very high, since polarisation is
an expected feature of a large number of physical phenomena of
astronomical interest. This is particularly true for unresolved
sources as GRB afterglows, where polarimetry offers one of the best
opportunity to infer on the real geometry of the system. In
particular, time resolved polarimetry can in principle give
fundamental hints on the jet luminosity structure and on the evolution
of the expanding fireball. This would provide reliable tools to
discriminate among different scenarios. Finally, it has been recently
realised that polarimetry of GRB afterglows can offer a direct way to
study the physical condition of the Inter-Stellar Medium (ISM) around
the GRB progenitor. GRB polarimetry, thus, becomes a powerful probe
for gas and dust in cosmological environments, a valuable research
field by itself.

In the following of this contribution we want to briefly comment on
the most recent advancement in the field and discuss the likely future
perspectives that are now open by the advent of the GRB dedicated
Swift satellite with its unprecedented rapid localisation capabilities
\citep{Geh04}.

\section{Synchrotron and beaming?}

The first pioneeristic attempts, culminated with the successful
observation of a $\sim 1.7$\% polarisation level in GRB\,990510
\citep{Cov99,Wij99}, were driven by the hypothesis that the afterglow
emission were due to synchrotron radiation \citep{Pac93, Mez97,
Sari98}. GRB\,990510 was also a perfect case for testing the
hypothesis of a geometrically beamed fireball.  Indeed, the detection
of an achromatic break in the optical light curve \citep{Isr99,
Harr99}, together with the observed degree of polarisation, gave
support to this scenario.  Shortly after this result, it was realised
that a jetted ultra-relativistic outflow would produce a characteristic
time evolution of the polarisation degree and position angle
\citep{Ghis99, Sar99}. The detailed shape of the polarisation
curves depends on the dynamical evolution.  Testing this model against
data is thus a powerful diagnosis of the geometry and dynamics of the
fireball.

A large number of polarimetric observations has been carried out since
GRB\,990510. A review of these data has been compiled by \citet{Cov04}
and \citet{Bjor03}. However, until recently, the detection of a low
level of polarisation required strong observational efforts. This
prevented a satisfactory time coverage of the afterglow decay and, in
turn, a convincing test for the model predictions.

\section{Homogeneous, Structured and Magnetised Jets}

\begin{figure}
\centering \includegraphics[width=0.75\textwidth,height=6cm]{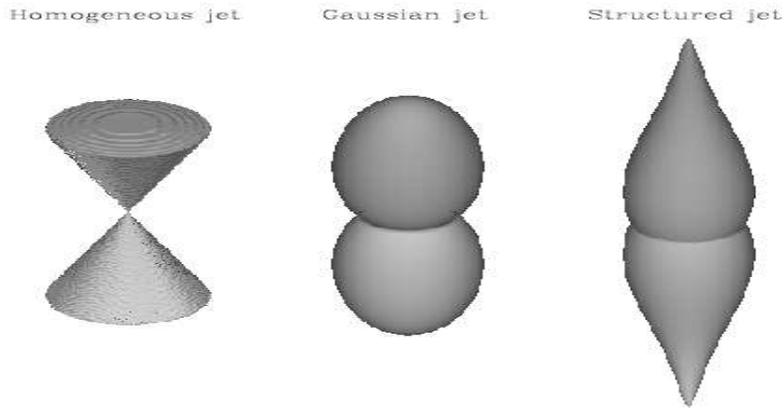} 
\caption{Possible different jet structures. From \citet{Ros04}.}
\label{fig:jets}
\end{figure}

\begin{figure}
\centering \includegraphics[width=\textwidth,height=10cm]{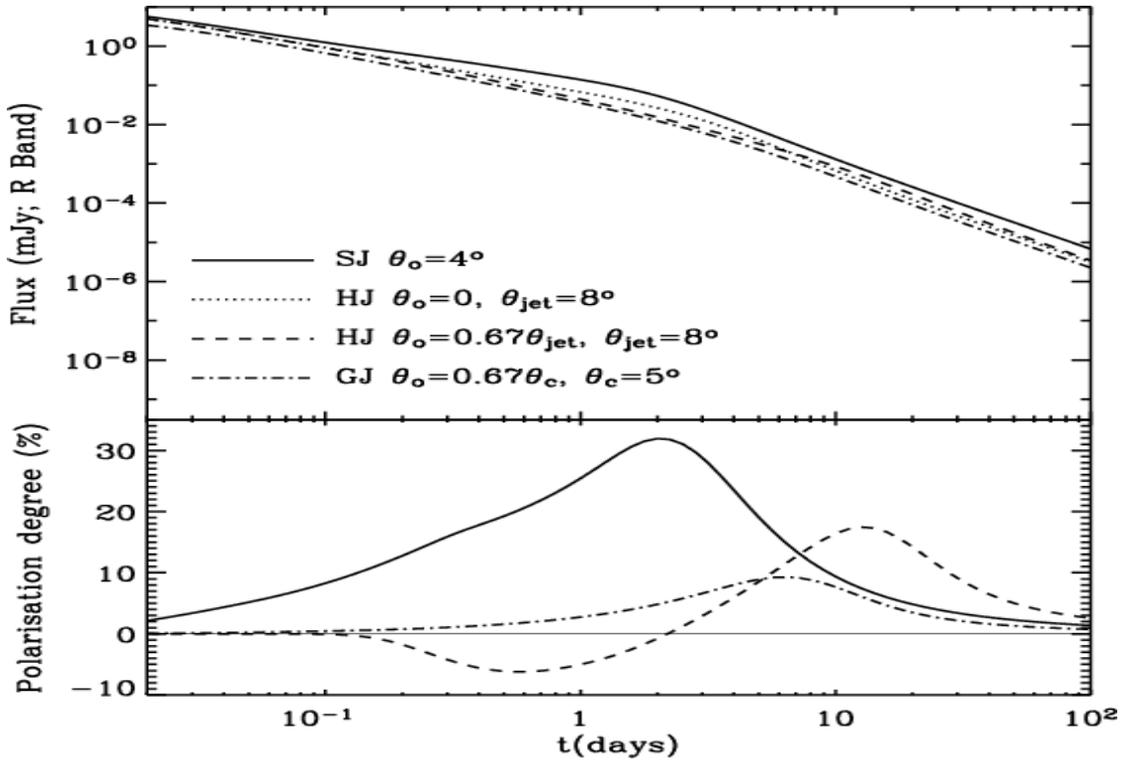}
\caption{Light curve and polarisation evolution for different jet 
structures. SJ stands for structured jet, HJ homogeneous jet, GJ for
Gaussian jet. The figure shows the similarity of the predicted light
curves for the various models while the polarisation changes
considerably. Negative polarisation degrees mark a 90$^\circ$ rotation
for the position angle. From \citet{Ros04}.}
\label{fig:polev}
\end{figure}

Lacking strong observational constraints, an improvement of the
reference models was achieved considering more physical descriptions
for the GRB afterglow jets. In the basic model the energy distribution
is homogeneous, making the jet a single entity. More complex beam and
magnetic field patterns (Fig.\,\ref{fig:jets}), reflecting a
physically more plausible scenario, were studied in several papers
\citep{Gran03,Laz04,Ros04} showing that the light curve is barely
affected by this parameter, while the polarisation and position angle
evolution changes substantially, providing a further diagnostic tool
Fig.~\ref{fig:polev}.

The universal structured jet model predicts that the maximum of the
polarisation curve is at the time of the break in the light curve. The
position angle remains constant throughout the afterglow evolution. On
the contrary, the homogeneous jet model requires two maxima before and
after the light curve break and, more importantly, the position angle
shows a sudden rotation of 90$^\circ$ between the two maxima, roughly
simultaneously to the break time of the light curve. At early and late
time the polarisation should be essentially zero
(Fig.\,\ref{fig:polev}).

This last result is substantially modified if it is assumed that a
large-scale magnetic field is driving the fireball expansion. The
topics has been widely discussed in the context of polarimetry by
\citet{Gran03}, \citet{Laz04} and \cite{Ros04}. Magnetised jets can be
both homogeneous and structured. We do not discuss here the details of
this recent research branch. However, we note that, at early times, a
large-scale ordered magnetic field produces a non negligible degree of
polarisation, contrary to the purely hydrodynamical models.
Polarimetry may therefore be the most powerful available diagnostic
tool to investigate the fireball energy content and its early
dynamical evolution.

\subsection{Dust Induced Polarisation}

\begin{figure}
\centering \includegraphics[width=\textwidth]{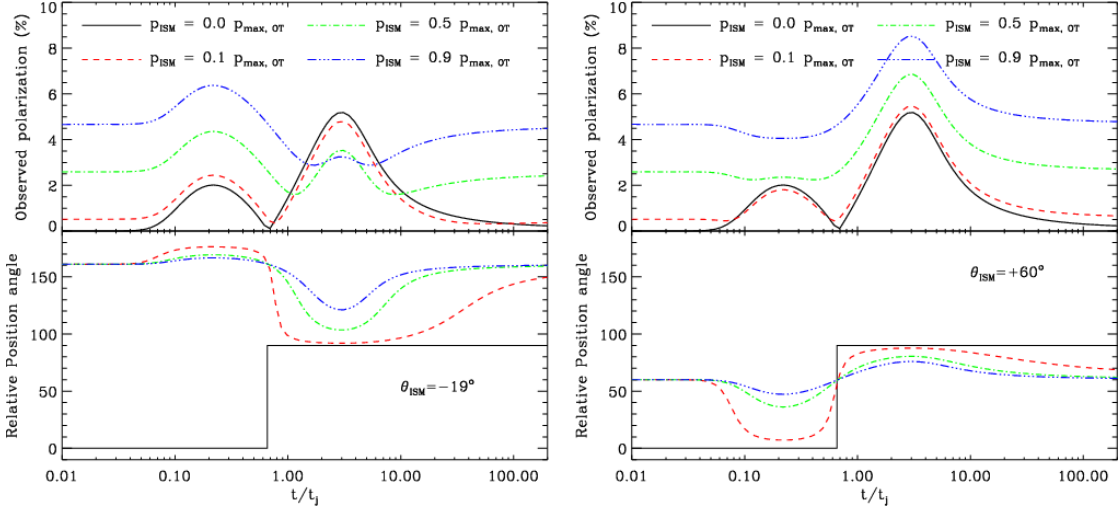}
\caption{Assuming as a reference a typical polarisation curve with 
a homogeneous jet, the presence of some dust along the line of sight
deeply modify the observed time evolution if the dust-induced
polarisation is comparable to the intrinsic one, as it seems to be the
rule for GRB afterglow at least at rather late time after the
high-energy event \citep{Cov04}. Depending on the relation between the
position angle of the dust-induced polarisation and of the intrinsic
GRB afterglow polarisation, the typical shape of the curve can be
removed or even enhanced. From \citet{Laz03}.}
\label{fig:dust}
\end{figure}

The observed low polarisation level from GRB afterglows is often
comparable to the expected polarisation induced by dust. Dust grains
are known to behave like a dichroic, possibly birefringent, medium
\citep{Laz03}. Significant amounts of dust are expected to lie close
to the GRB site, as a consequence of the observation of a supernova
(SN) component in a few GRBs.  The measured polarisation will be
modified by the propagation of radiation through dusty media. This
effect is, contrary to the intrinsic afterglow polarisation,
wavelength dependent. The different wavelength dependence open the
interesting possibility to study the polarisation signature from the
afterglow to study the physical characteristics of dust in
cosmological environments: probably the only way to study dust close
to star formation regions at high redshift. Even assuming that dust
properties close to GRB formation sites are comparable to what we know
in the Milky Way (MW), it is important to take into account this
component once information from time evolution polarimetry are
derived. The superposition of the intrinsic time evolution to
dust-induced components for the GRB host galaxy and the MW may
substantially alter the expected behavior (Fig.\,\ref{fig:dust}).

\section{Observations vs. Theory}

So far, a rather satisfactory coverage of the polarisation evolution
of a GRB afterglow has been obtained for three events only:
GRB\,021004 \citep{Rol03,Laz03,Nak04,Bjor04}, GRB\,030329
\citep{Gre03,Klo04}, and GRB\,020813 \cite{Goro04,Laz04}. However,
firm conclusions from the analysis could have been derived for the
last case only. GRB\,021004 and GRB\,030329 showed some remarkable
similarities given that their light curves were characterised by a
large number of ``bumps'' or rebrightenings. Several different
possibilities has been proposed to model the irregularities in the
light curve invoking clumping in the external medium \citep{Laz02}; a
more complex and not axi-symmetric energy distribution in the fireball
\citep{Nak04} or delayed energy injections \citep{Bjor04}. It was soon
clear \citep{Laz03} that the standard models for polarisation could
not be applied in these conditions, since they are all derived in
cylindrical symmetry.  Even for GRB\,030329, for which a remarkable
dataset was obtained \citep{Gre03}, no convincing explanation of the
polarization and light-curve erratic behaviors has so far been
obtained.  It is not clear yet to what extent GRB\,021004 and
GRB\,030329 belong to the same population of long GRBs. It is argued
however that the failed detection of this erratic behavior in other
afterglows (such as GRB~020813) is not due to a coarser sampling of
the light curve.

\begin{figure}
\centering \includegraphics[width=\textwidth,height=8cm]{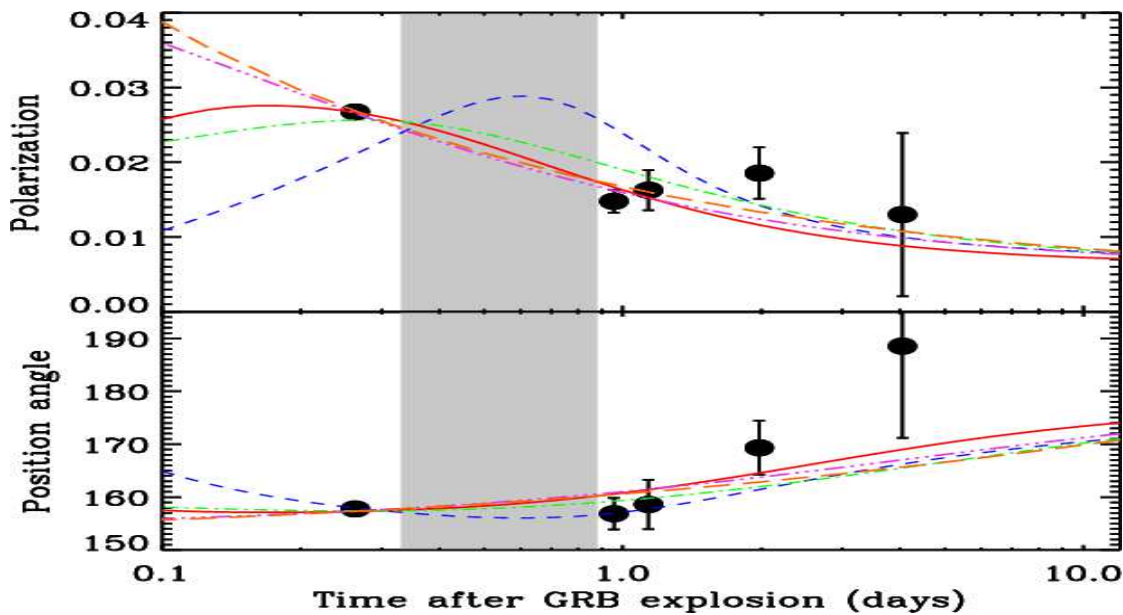}
\caption{Polarisation data for GRB\,020813 \citep{Goro04}. 
Different curves refer to different models. From \citet{Laz04}.}
\label{fig:020813}
\end{figure}

GRB\,020813 was the best case for model testing. Its light curve was
remarkably smooth \citep{Cov03}, in several optical/infrared bands,
and a break in the light curve was clearly singled out. A few
polarimetric observations have been carried out providing for the
first time polarisation data before and after the light curve break
time \citep{Goro04}. \citet{Laz04} applied to this event a more
quantitative approach not limited, as usually done in the past, to the
bare qualitative search of features in the polarisation curve
(i.e. rotation of the position angle, etc.). A formal analysis was
carried out, taking into account the GRB host galaxy and MW dust
induced polarisation and the intrinsic GRB afterglow polarisation.
All current jet models were considered, including homogeneous and
structured jets, with and without a coherent magnetic field. The
dataset, did not allow us to strictly derive a best fitting model.
The main result was to rule out the basic homogeneous jets model at a
confidence larger than $3\sigma$, mainly because of the lack of the
predicted $90^\circ$ position angle rotation. Again the role of the MW
dust induced polarisation is significant. All magnetized models and
structured jets fit satisfactorily the data, the ambiguity being
mainly due to the lack of early time measurement, i.e. where
magnetised or not magnetised models mostly differ
(see Fig.\,\ref{fig:020813}).

The debate is still far from being settled. Recently, for GRB\,030226
\citet{Klo04b} a quite low upper limits ($\sim 1$\%) was reported, in
rather strict coincidence with the break time, therefore close to the
maximum for the polarisation curve if we assume a structured jet
model. With one only measurement it is difficult to draw firm
conclusions, since this null polarisation measurement may well be due
to dust induced polarisation superposed destructively to the
intrinsic, if any, GRB afterglow polarisation.

It is finally worth, even though tautological, to report that, as soon
as Swift will be fully operational, distributing routinely prompt
localisations, a new era will be open even for GRB polarimetry.  It
will allow us to carry out more stringent tests to the available
models and therefore strictly constraint geometry, energetics and
dynamics of the fireball.


\begin{thebibliography}{9}
\bibitem[Gehrels et al. (2004)]{Geh04} Gehrels, N., Chincarini, G., 
Giommi, P., et al. 2004, ApJ 611, 1005
\bibitem[Covino et al. (1999)]{Cov99} Covino S., Lazzati D., Ghisellini G., 
et al. 1999, A\&A 348, 1
\bibitem[Wijers et al. (1999)]{Wij99} Wijers R.A.M.J., Vreeswijk P.M., 
Galama T.J., et al. 1999, ApJ 523, 177 
\bibitem[Paczy\'nski \& Rhoads (1993)]{Pac93} Paczy\'nski B., Rhoads J.E. 
1993, ApJ 418, 5
\bibitem[M\'esz\'aros \& Rees (1997)]{Mez97} M\'esz\'aros P., Rees M.J. 
1997, ApJ 476, 232
\bibitem[Sari et al. (1998)]{Sari98} Sari R., Piran T., Narayan R. 1998, 
ApJ 497, 17
\bibitem[Israel et al. (1999)]{Isr99} Israel G.L., Marconi G., Covino S., 
et al. (1999), A\&A 348, 5
\bibitem[Harrison et al. (1999)]{Harr99} Harrison  F.A., Bloom J.S., 
Frail D.A., et al. (1999), ApJ 523, 121
\bibitem[Ghisellini \& Lazzati (1999)]{Ghis99} Ghisellini G., Lazzati D. 
(1999), MNRAS 309, 7
\bibitem[Sari (1999)]{Sar99} Sari R. (1999), ApJ 524, 43
\bibitem[Covino et al. (2004)]{Cov04} Covino S., Ghisellini G., Lazzati D., 
Malesani D. 2004, ASP Conf. Ser. 312, 169
\bibitem[Bj\"ornsson (2003)]{Bjor03} Bj\"ornsson G. (2003), astro-ph/0302177  
\bibitem[Granot \&  K\"onigl (2003)]{Gran03} Granot J., K\"onigl A. (2003), 
ApJ 594, 83
\bibitem[Lazzati et al. (2004)]{Laz04} Lazzati D., Covino S., Gorosabel J.R., 
et al. (2004), A\&A 422, 121
\bibitem[Rossi et al. (2004)]{Ros04} Rossi E.M., Lazzati D., Salmonson J.D., 
Ghisellini G. (2004), MNRAS 354, 86
\bibitem[Lazzati et al. (2003)]{Laz03} Lazzati D., Covino S., di Serego 
Alighieri S., et al. (2003), A\&A 410, 823
\bibitem[Rol et al. (2003)]{Rol03} Rol E., Wijers R.A.M.J., Fynbo J.P.U. 
et al. (2003), A\&A 405, 23
\bibitem[Nakar \& Oren (2004)]{Nak04} Nakar E., Oren Y. (2004), ApJ 602, 97
\bibitem[Bj\"ornsson et al. (2004)]{Bjor04} Bj\"ornsson G., 
Gudmundsson E.H.,  J\'ohannesson G. (2004), ApJ 615, 77
\bibitem[Greiner et al. (2003)]{Gre03} Greiner J., ???, et al. (2003), 
Nature 426, 157
\bibitem[Klose et al. (2004)]{Klo04} Klose S., Palazzi E., Masetti N., 
et al. (2004),  A\&A 420, 899
\bibitem[Gorosabel et al. (2004)]{Goro04} Gorosabel J., Rol E., Covino S., 
et al. (2004), A\&A 422, 113
\bibitem[Lazzati et al. (2002)]{Laz02} Lazzati D., Rossi E., Covino S., 
Ghisellini G., Malesani D. (2002), A\&A 395, 5
\bibitem[Covino et al. (2003)]{Cov03} Covino S., Malesani D., Tavecchio F. 
et al. (2003), A\&A 404, 5
\bibitem[Klose et al. (2004b)]{Klo04b} Klose S., Greiner J., Rau A. 
et al. (2004b), AJ 128, 1942
\end{thebibliography}
\end{document}